\documentstyle[prd,aps,twocolumn]{revtex}
\tighten

\begin{document}
\draft
 
\pagestyle{empty}

\preprint{
\noindent
%\begin{minipage}[t]{3in}
%\begin{flushleft}
%\today \\
%\end{flushleft}
%\end{minipage}
\hfill
\begin{minipage}[t]{3in}
\begin{flushright}
LBL--49711 \\
UCB--PTH--02/10 \\
hep-ph/0203012 \\
%month 1996
\end{flushright}
\end{minipage}
}

\title{Elusive vector glueball}

\author{
Mahiko Suzuki
%\thanks{Work supported in part by the Director, Office of Energy
%Research, Office of High Energy and Nuclear Physics, Division of High
%Energy Physics of the U.S. Department of Energy under Contract
%DE--AC03--76SF00098 and in part by the National Science Foundation under
%grant PHY--00--xxxxx.}
}
\address{
Department of Physics and Lawrence Berkeley National Laboratory\\
University of California, Berkeley, California 94720
}

%\thanks{Work supported by the Department of Energy under Contract
%DE--AC03--76SF00515.}

\date{\today}
\maketitle

\begin{abstract}
      If the vector glueball ${\cal O}$ exists in the mass range 
that theory suggests, its resonant production cross section can be
detected in $e^+e^-$ annihilation only if the decay width is very 
narrow ($\leq$ a few MeV). Otherwise ${\cal O}$
will be observed only indirectly through its mixing with 
$\psi'$. We propose a few tests of the ${\cal O}$-$\psi'$ 
mixing for future charm factories.   
  
\end{abstract}

\pagestyle{plain}
\narrowtext
\setcounter{footnote}{0}

\section{Introduction}

    While theorists expect that the glueballs exist as the 
flavorless hadrons made of gluons alone, the predicted masses 
vary from lattice QCD\cite{Lattice} to phenomenological 
models such as the bag-type model\cite{bag} and a naive constituent 
model\cite{Hou}. Experiment provided a few candidates for 
the ground-state glueball of $J^{PC}=0^{++}$ in the past, but no 
consensus has been reached. The glueball ${\cal O}$ of $1^{--}$ 
is made of three gluons in the constituent picture so that its mass 
is expected to be roughly 50\% higher than the lowest glueball mass. 
The recent lattice QCD calculation\cite{Lattice} predicts the mass 
of ${\cal O}$ at $3850 \pm 50 \pm 190$ MeV in the quenched 
approximation. It is amusing that this value happens to be very 
close to the $\psi'$ mass (3686 MeV). In contrast to the masses, 
very little is known even theoretically about the widths since 
they are determined by how the glueballs couple to light quarks. 

     A few puzzles exist in the hadronic $\psi'$ decay.
One is the so-called $\rho\pi$ puzzle, {\em i.e.,} the severe 
suppression of $\psi'\to\rho\pi$\cite{Franklin}. If both $J/\psi$ 
and $\psi'$ decay into hadrons through the perturbative $ggg$ 
from the $^3S_1$ states of $c\overline{c}$, we expect 
that their hadronic decays should be very similar to each other. 
Contrary to this expectation, the branching fractions are very 
different not only for $\rho\pi$ but also for many other exclusive 
hadron channels\cite{Harris}. Since ${\rm B}(J/\psi\to\rho\pi)/{
\rm B}(J/\psi\to\omega\pi)$ is close to ${\rm B}(J/\psi\to ggg\to 
hadrons)/{\rm B}(J/\psi\to\gamma^*\to hadrons)$\cite{PDG}, 
the decay $J/\psi\to\rho\pi$ is just as strong as the naive 
expectation. The experimental fact that 
${\rm B}(\psi'\to\omega\pi)/{\rm B}(J/\psi\to\omega\pi)\approx 
{\rm B}(\psi'\to\ell^+\ell^-)/{\rm B}(J/\psi\to\ell^+\ell^-)$ 
assures us that there is nothing anomalous about the $1^-0^-$
modes in the one-photon annihilation. There appears to be no 
perturbative $1^-0^-$ suppression that was once suggested as 
a resolution to the $\rho\pi$ puzzle\cite{Brodsky}. Therefore
the source of the $\rho\pi$ puzzle is in some property of $\psi'$.

 The other puzzle is an apparent excess of $\psi'\to hadrons$ 
relative to $J/\psi\to hadrons$\cite{Suzuki,Chinese}. To establish 
it quantitatively, we need more an accurate determination of the 
cascade decay branchings of $\psi'$. The current data hint that 
there exists some unknown decay mechanism for $\psi'\to hadrons$ 
other than the standard perturbative three-gluon decay. It is the 
$\psi'$-${\cal O}$ mixing that may resolve both puzzles by one 
shot\cite{Suzuki}.\footnote{
The other idea is a large virtual $D\overline{D}$ configuration
or a large $^3D_1$ mixing in the $\psi'$ state\cite{Rosner}.
All earlier attempts to resolve the $\rho\pi$ puzzle failed with 
the recent analysis of the $\psi'$ decay modes, as was shown by 
Harris\cite{Harris}. See also Ref.\cite{Tuan}.}

 In view of the precision charm experiment proposed at CESR 
and at Beijing, we pursue here the ${\cal O}$-$\psi'$ 
mixing as quantitatively as possible. Assuming 
that the excess $\psi'\to hadrons$ really exists and that it is 
caused by the ${\cal O}$-$\psi'$ mixing, we estimate the resonant 
${\cal O}$ production cross sections. We find that the ${\cal O}$ 
resonance peak will be detected in $e^+e^-$ annihilation only if 
the decay width is very narrow ($\Gamma_{{\cal O}} <$ a few MeV). 
If the width is wider, we shall be able to find only indirect 
evidences for ${\cal O}$ through the $\psi'$ decay. 
In a dedicated charmonium experiment, experimentalists should 
first confirm the excess in $\psi'\to hadrons$ by determining 
more accurately the $\psi'\to J/\psi$ cascade decay branchings, and
then examine individual exclusive hadron channels of even G-parity 
such as $\omega\pi$, $\pi^+\pi^-$, and $\rho\eta$.

\section{${\cal O}$-$\psi'$ mixing}

    The ${\cal O}$-$\psi'$ mixing is determined by the $2\times 2$ 
mass matrix of ${\cal O}$ and $\psi'$, 
\begin{equation}
   {\cal M}  =  \left( \begin{array}{cc}
   m^2_{{\cal O}}-im_{{\cal O}}\Gamma_{{\cal O}} & f_{{\cal O}\psi'} \\
   f_{{\cal O}\psi'} & m^2_{\psi'}-im_{\psi'}\Gamma_{\psi'}
               \end{array} \right)  \label{mass}.
\end{equation}
Two eigenstates of ${\cal M}$ are the standard Breit-Wigner resonances. 
Since the complex mixing angle $\theta\simeq\frac{1}{2}\tan^{-1}
[2f_{{\cal O}\psi'}/({\cal M}_{11}-{\cal M}_{22})]$ is very small 
in magnitude in the cases of our interest, we shall treat the mixing 
perturbatively and denote the physical Breit-Wigner states also by 
${\cal O}$ and $\psi'$ below. We shall discard 
$|m_{\psi'}-m_{{\cal O}}|/m_{\psi'}\ll 1$ wherever appropriate.

   In addition to the standard perturbative $ggg$ decay, the physical 
$\psi'$ state can decay through its mixing with ${\cal O}$:
\begin{equation}
   \Gamma(\psi'\to{\cal O}\to all)
  = \frac{|f_{{\cal O}\psi'}|^2}{m_{{\cal O}}^2\Gamma_{{\cal O}}}
     F(\Delta m,\Gamma). \label{rate}
\end{equation}
where $F(\Delta m,\Gamma)$ comes from the ${\cal O}$ 
resonance shape;
\begin{equation}
        F(\Delta m,\Gamma) = \Gamma_{{\cal O}}^2/[4(\Delta m)^2+
             \Delta\Gamma_{{\cal O}}^2] \label{F}
\end{equation} 
with $\Delta m = |m_{{\cal O}}-m_{\psi'}|$ and $\Delta
\Gamma_{{\cal O}}\simeq\Gamma_{{\cal O}} (\gg\Gamma_{\psi'})$.

   We parametrize the alleged excess in the hadronic $\psi'$ decay as
\begin{equation}
   \Gamma(\psi'\to hadrons)/\Gamma(\psi'\to ggg) = 1 + {\cal E},
\end{equation}
where ${\cal E}\approx$ 0.6-0.7 with large errors according to 
the current data\cite{Suzuki,Chinese}. This excess is attributed 
to the additional decay, $\psi'\to{\cal O}\to hadrons$.
Since $\Gamma_{{\cal O}}\gg \Gamma_{\psi'}$, even a tiny 
mixing can make this contribution as large as the 
perturbative $\Gamma(\psi'\to ggg)$. Without knowing dynamics of 
${\cal O}$ decay, let us assume for the moment that $\psi'\to {\cal O}
\to hadrons$ adds to $\psi'\to ggg\to hadrons$ in the total 
rate. That is, we assume no interference in the inclusive decay
rate, though the interference is crucial in individual exclusive rates. 
This is not an absurd assumption: If the $ggg$ state of ${\cal O}$ 
differs substantially from the perturbative $ggg$ configuration
of $\psi'$ decay, it is capable of producing very different interference 
patterns for one exclusive channel to another, leading to suppression 
or enhancement of the exclusive rates, yet the interference terms cancel 
out in the inclusive rate. On this assumption, $f_{{\cal O}\psi'}$ is 
constrained by the excess ${\cal E}$ as
\begin{equation}
   |f_{{\cal O}\psi'}|^2 = 
   \frac{(25 {\rm MeV})^4{\cal E}}{F(\Delta m,\Gamma)}
   \times\biggl(\frac{\Gamma_{{\cal O}}}{1\;{\rm MeV}}\biggr). \label{f}
\end{equation}
The $\psi'\to\gamma^*$ transition strength defined by $f_{\gamma\psi'}
A^{\mu}\psi'_{\mu}$ can be expressed as 
$f_{\gamma\psi'}^2=(3m_{\psi'}^3/\alpha)\Gamma(\psi'\to e^+e^-)$.

{\em If} ${\cal O}$ decays mainly as ${\cal O}\to q\overline{q}
\to hadrons$, and {\em if} the couplings of ${\cal O}$ to quarks 
are flavor blind, we could derive an additional interesting constraint 
with Eq. (\ref{f}) by expressing $f_{{\cal O}\psi'}$ with the 
${\cal O}c\overline{c}$ coupling, and the ${\cal O}q\overline{q}$ 
coupling with $\Gamma_{{\cal O}}$. 
The result is $\Delta m/\Gamma_{{\cal O}}\simeq 30/\sqrt{{\cal E}}$.

\section{Resonant production cross sections}

\subsection{$e^+e^-\to{\cal O}\to all$}

     The glueball ${\cal O}$ is most favorably reached 
through $e^+e^-\to \gamma^* \to \psi' \to {\cal O}\to all$.  
The cross section at the ${\cal O}$ resonance peak is given by
\begin{equation}
  \sigma(e^+e^-\to{\cal O}\to all) =
     \frac{\pi\alpha|f_{\gamma\psi'}f_{{\cal O}\psi'}|^2}{
     m_{{\cal O}}^7(\Delta m)^2\Gamma_{{\cal O}}}.
\end{equation}
In the unit of $\sigma(e^+e^-\to\mu^+\mu^-)$, the peak value is
\begin{equation}
    R = \frac{9|f_{{\cal O}\psi'}|^2\Gamma(\psi'\to\ell^+\ell^-)}{
        4\alpha^2(\Delta m)^2 m_{{\cal O}}^2 \Gamma_{{\cal O}}}.
\end{equation}
The right-hand side is sizable only if the width $\Gamma_{{\cal O}}$ 
is very narrow and $|f_{{\cal O}\psi'}|^2/\Gamma_{{\cal O}}$ is 
strongly enhanced by small $F(\Delta m,\Gamma)$. (See Eq. (\ref{f}).)
For $\Gamma_{{\cal O}}\ll\Delta m$, $R$ is insensitive to $\Delta m$ 
and scales approximately as $1/\Gamma_{\cal O}^2$;
\begin{equation}
    R\simeq 
  \biggl(\frac{3.2\;{\rm MeV}}{\Gamma_{{\cal O}}}\biggr)^2{\cal E}.
\end{equation}
If ${\cal O}$ decays with three gluons falling apart into quarks
by long-distance interactions,
such a narrow width is unlikely. On the other hand, if short-distance 
physics suppresses conversion of $ggg$ into light quarks by 
$\alpha_s^3$, a very narrow width is not totally out of question.
Note that $\Gamma=0.087$ MeV for $J/\psi$ by $\alpha_s^3$ 
suppression. Since we have no knowledge of how the glueballs decay, 
there is no orthodoxy about their widths. 

 Suppose that one finds a candidate of the ${\cal O}$ peak by 
scanning over the relevant energy range. If it is really the 
${\cal O}$ resonance, the lepton-pair production cross section 
remains the same at the peak as at the side bands:
\begin{equation}
    R_{\ell^+\ell^-} = 1, \;\;(peak),
\end{equation}  
since ${\cal O}\to q\overline{q}\to\gamma^*\to\ell^+\ell^-$ is 
negligible. Another test is whether the increase in hadron events 
occurs all in $I=0$ or not. This is most easily checked by $G$-parity 
of the final states. At the continuum off the resonance, 
hadrons are produced through light quarks in $I=0$ and $I=1$ channels 
in the ratio of
\begin{equation}
    R_{I=0}/R_{I=1} = 1/3,\;\; (continuum). \label{G}
\end{equation}
That is, if one counts the final states of even pions and of odd 
pions at the continuum, the even-pion events dominate over the 
odd-pion events. This odd-to-even pion ratio should 
suddenly jump at the ${\cal O}$ resonance.  Narrowness of width 
rules out light-quark resonances of $I=0$.

 We comment on production of ${\cal O}$ through the light quarks 
instead of $\psi'$: $e^+e^-\to q\overline{q}(q=u,d,s)\to{\cal O}
\to hadrons$. This production process is actually 
flavor-SU(3)-forbidden since the electromagnetic current of the 
$u,d,s$ quarks is an octet while ${\cal O}$ is a singlet. 
Production of ${\cal O}$ occurs only by SU(3) breakings. We can 
compute these amplitudes perturbatively, if we keep only the 
on-shell contribution of $q\overline{q}$. The result is independent 
of $\Gamma_{{\cal O}}$ if $\Gamma_{{\cal O}}$ is given by the
process  ${\cal O}\to q\overline{q}\to hadrons$:
\begin{equation}
 R = \frac{1}{48}\biggl(\frac{2m_s}{m_{{\cal O}}}\biggr)^4, 
                         \label{light}
\end{equation} 
where $m_s$ is the $s$-quark mass that enters through the quark 
mass splitting of the intermediate states. 
Even if ${\cal O}$ is produced through $s\overline{s}$ alone 
or through $u\overline{u}+d\overline{d}$ by the maximal SU(3) 
violation, the value of $R$ would be only 0.037. The off-shell 
$q\overline{q}$ contribution depends on the form factor damping 
of the ${\cal O}q\overline{q}$ vertex, which is expected to be 
soft since ${\cal O}$ is an extended object. Therefore the 
off-shell contribution cannot be much larger than the on-shell 
contribution. We may safely dismiss the ${\cal O}$ 
production through light quarks.

\subsection{$p\overline{p}\to{\cal O}\to hadrons$}

    The coupling of ${\cal O}$ to $p\overline{p}$ is allowed by 
flavor SU(3). However, it is most likely small because so many 
channels are open aside from $p\overline{p}$ when $ggg$
of ${\cal O}$ turn in hadrons. Although accurate estimate of 
this coupling is not possible, we can make a reasonable guess  
with the charmonium decay branching into $p\overline{p}$.

  The $J=1$ partial-wave cross section for $p\overline{p}\to{\cal O}
\to all$ in the resonance region is given by the standard formula,
\begin{equation}
 \sigma_{res}(p\overline{p}\to{\cal O}\to all)
  =\frac{3\pi}{|{\bf p}|^2}
 \frac{B_{in}\Gamma_{{\cal O}}^2}{4(W-m_{{\cal O}})^2+\Gamma_{{\cal O}}^2},
\end{equation}
where $W$ and ${\bf p}$ are the {\em cm} energy and momentum, 
and $B_{in}$ is the decay branching 
fraction for ${\cal O}\to p\overline{p}$. Since $J/\psi$
decays through $ggg$, a reasonable guess is to equate $B_{in}$ with 
${\rm B}(J/\psi\to p\overline{p})/{\rm B}(J/\psi\to ggg)=0.24\%$.  
Then the peak cross section is
\begin{equation}
  \sigma_{peak} = 3.5 \mu{\rm b} 
\end{equation}
for $m_{\cal O}\simeq 3.7$ GeV. Since the $p\overline{p}$ total 
cross section is approximately 60 mb at this energy, there is no 
chance to see ${\cal O}$ as a resonance in $p\overline{p}$ annihilation.

Nonresonant production of ${\cal O}$ in conjunction with other light
hadron(s), {\em e.g.,} $p\overline{p}\to{\cal O}X$, is  suppressed 
by the feeble coupling of $q\overline{q}\to ggg$ (the ``disconnected
quark diagram'') for an identifiably narrow ${\cal O}$. This suppression 
is severe even for $\phi$, which is much lighter: $\sigma(\pi p\to 
\phi X)/\sigma(\pi p\to \omega X)\approx 1/100$. Unlike $J/\psi$ search
by $J/\psi\to\ell^+\ell^-$, absence of a discriminating decay 
characteristic will make it hard to identify ${\cal O}$.  

\section{Signals of ${\cal O}$-$\psi'$ mixing in $\psi'$ decay}

 The only chance to produce and directly observe the vector 
glueball ${\cal O}$ will be in the case that the decay width 
is very narrow. Our numerical estimate is based on the assumption 
that $\psi\to{\cal O}\to hadrons$ and $\psi'\to ggg \to hadrons$ 
add up without interference in the total rates. Most generally, 
${\cal E}$ should be left as a parameter of $O(1)$. If the width 
$\Gamma_{{\cal O}}$ happens to be wider than a few MeV 
and the resonance search is not feasible, ${\cal O}$ must be searched 
by the indirect signatures in the $\psi'$ decay that result from the 
${\cal O}$-$\psi'$ mixing. Let us explore such indirect signatures.

If $\psi'$ is purely an $s$-wave bound state of $c\overline{c}$, and 
its hadronic decay occurs through the perturbative three-gluon decay 
alone, we expect 
\begin{eqnarray}
 \frac{{\rm B}(\psi'\to ggg\to all)}{{\rm B}(J/\psi\to ggg\to all)} 
 &\simeq&\biggl(\frac{\alpha_s(m_{\psi'})}{\alpha_s(m_{J/\psi})}\biggr)^3
 \frac{{\rm B}(\psi'\to\ell^+\ell^-)}{{\rm B}(J/\psi\to\ell^+\ell^-)}.
  \nonumber \\
 & = & 0.13 \pm 0.03 \label{rule1}
\end{eqnarray}
The similarity of $\psi'$ 
to $J/\psi$ led us to suspect that the same relation may hold 
approximately for individual exclusive channels as well. This naive 
rule of thumb was referred to as ``the 14\% rule''. However, 
experiment has shown very large deviations from the ``rule'' 
for many decay modes\cite{comment}. The ratio is most prominently 
off for $\rho\pi$, by a factor of 60 or more. This is the 
$\rho\pi$ puzzle.

\subsection{Even-$G$ channels}     

    The ${\cal O}$-$\psi'$ mixing affects only the final 
states of $I=0$. Since $J^{PC}=1^{--}$, they are odd in G-parity.
Since the even-$G$ states fed by $\psi'\to\gamma^*\to hadrons$ 
are immune to the mixing, we should find
\begin{eqnarray}
 \frac{{\rm B}(\psi'\to h(G=+))}{{\rm B}(J/\psi\to h(G=+))} &\simeq&
  \frac{{\rm B}(\psi'\to\ell^+\ell^-)}{{\rm B}(J/\psi\to\ell^+\ell^-)}
  \nonumber\\
 &=& 0.15 \pm 0.03. \label{rule2}
\end{eqnarray}
    Among the decay modes of even $G$ so far analyzed, 
\begin{equation}
 \frac{{\rm B}(\psi'\to\omega\pi)}{{\rm B}(J/\psi\to\omega\pi)}
  = 0.09 \pm 0.05,
\end{equation} 
which is off from Eq. (\ref{rule2}), but only one standard 
deviation. A higher precision is desired for $\psi'\to\omega\pi$.
In comparison the $\rho\pi$ modes is severely suppressed as
\begin{equation}
  \frac{{\rm B}(\psi'\to\rho\pi)}{{\rm B}(J/\psi\to\rho\pi)}
  < 0.002.
\end{equation}     
The same ratio is $<$0.021 for $\omega\eta$ and $\rho a_2$, which are 
both odd in $G$. It is not true that all odd-$G$ channels are suppressed; 
the ratio is $0.18 \pm 0.05$ for $b_1^{\pm}\pi^{\mp}$ over 0.13 
of Eq. (\ref{rule1}). The ratio is much larger for $\omega\eta'$
($0.5\pm 0.4$), though the error is large.

  There are a few other even-$G$ channels in the $J/\psi$ decay for 
which good measurements already exist. Among them are $\pi^+\pi^-$, 
$\rho\eta$, and $\rho\eta'$. These decay modes should be measured
carefully for $\psi'$ for comparison. There is one $\psi'$ decay 
mode that was observed with a large branching fraction but not yet 
identified in the $J/\psi$ decay. That is, $\psi'\to\rho^0 \pi^+\pi^-$ 
with ${\rm B}(\psi'\to\rho^0\pi^+\pi^-) = (3.7 \pm 0.6 \pm 0.9)\times 
10^{-4}$\cite{Harris}. Only ${\rm B}(J/\psi\to\pi^+\pi^-\pi^+\pi^-) 
= (4.0\pm 1.0)\times 10^{-3}$\cite{PDG} is quoted for $J/\psi$\cite{PDG}. 
If the ${\cal O}$-$\psi'$ mixing is the solution to the puzzles, 
${\rm B}(J/\psi\to\rho^0\pi^+\pi^-)$ should obey Eq. (\ref{rule2})
so that a large fraction of $J/\psi\to\pi^+\pi^-\pi^+\pi^-$ 
should consist of $J/\psi\to\rho^0\pi^+\pi^-$. No attempt has yet 
been made to search $a_J\pi$ (J=0,1,2) in $\pi^+\pi^-\pi^+\pi^-$. 
It is worth efforts for testing of the ${\cal O}$-$\psi'$ mixing.

\subsection{$\gamma h$ modes}

   Some final states that consist of one photon and a hadron have
large branching fractions. It is often interpreted that they arise
from $J/\psi(\psi')\to\gamma gg\to \gamma h$ since the $\alpha$
suppression is partially compensated with lack of one power of $\alpha_s$ 
in $\gamma gg$. If this interpretation is right, the ${\cal O}$-$\psi'$ 
mixing should little affect the $\gamma h$ modes. Then we would expect
\begin{equation}
   \frac{{\rm B}(\psi'\to h\gamma)}{{\rm B}(J/\psi\to h\gamma)}\simeq 
   \frac{{\rm B}(\psi'\to\ell^+\ell^-)}{{\rm B}(J/\psi\to\ell^+\ell^-)}.
\end{equation}
At present, three decay modes are available for comparison: The ratios 
are $0.06\pm 0.04$ for $\gamma\eta$, $0.036\pm 0.012$ for $\gamma\eta'$, 
and $0.22\pm 0.16$ for $\gamma f_2(1270)$. The errors are still large for 
$\gamma\eta$ and $\gamma f_2$.  But the ratio for $\gamma\eta'$
is clearly off the ${\cal O}$-$\psi'$ mixing predicton, $0.15\pm 0.03$. 
This may be taken against the ${\cal O}$-$\psi'$ mixing. Or else the 
decay branching ${\cal O}\to\gamma\eta'$ is enhanced for some reason. 
We recall that the $\eta'$ yield tends to be larger than theoretical 
expectations in many other processes, {\em e.g.}, $B$ decay. The common 
theme is the QCD anomalies involved in them.  Together with higher 
precision in experiment, better theoretical understanding is needed 
for $\gamma h$, particularly for $\gamma\eta'$.    

\subsection{Alternatives to ${\cal O}$-$\psi'$ mixing}

   An alternative to the ${\cal O}$-$\psi'$ mixing is to postulate 
that $\psi'$ contains a large $d$-wave component of $c\overline{c}$, 
much larger than we have so far accepted. The $d$-wave content of 
$\psi'$ was studied in the nonrelativistic model in the 
past\cite{KY,Banff}. The amount of the $d$-wave appears to be 
considerably smaller than what is relevant to the issues 
in the $\psi'$ decay. There is also an inconsistency with 
the radiative decays of $\psi'$\cite{Rosner}. However, these 
problems may well be due to the nonrelativistic potential model. 
Let us abandon the nonrelativistic model for the moment and assume 
in an ad hoc manner that a large $d$-wave component exists in $\psi'$.
In this case $\psi'\to \ell^+\ell^-$ would also be affected by the
mixing, but differently from $\psi'\to ggg \to hadrons$. Therefore, 
the branching fractions of even-$G$ channels would no longer 
obey the rule of Eq. (\ref{rule2}) precisely. 

   One variation of this scenario is to replace the large $d$-wave
component with the virtual $D\overline{D}$ configuration. In
this case $D\overline{D}\to \ell^+\ell^-$ is dynamically suppressed 
since both $c\overline{c}$ and $q\overline{q}$ ($q=u,d$) must be 
annihilated into $\gamma^*$ to create $\ell^+\ell^-$. In other words, 
the form factor suppression of $D\overline{D}$ occurs relative to 
the hard $c\overline{c}$ annihilation. Therefore $\psi'\to
\ell^+\ell^-$ is reduced by the amount of the $D\overline{D}$ 
component. Then Eq. (\ref{rule1}) would no longer hold. Although 
the hadrons from $D\overline{D}$ ($D=c\overline{u}, c\overline{d}$) 
contains an SU(3)-octet $I=0$ component ($\lambda_8$) in addition 
to an SU(3) singlet, it will not be easy to identify this octet 
mixture, separating from the isoscalar electromagnetic contribution 
and the strong SU(3) breaking effects.  

\section{Summary and remarks}

 The chance of finding ${\cal O}$ exists only if its width is 
very narrow by the short-distance QCD suppression. Otherwise 
${\cal O}$ will show up only in the $\psi'$ decay indirectly. 
Even if the $\rho\pi$ puzzle is set aside, the proximity of 
the predicted ${\cal O}$ mass to the $\psi'$ mass inevitably 
generates some ${\cal O}$-$\psi'$ mixing. This will be the only
experimental clue to the vector glueball of width wider than
a few MeV. It is therefore worth studying the mixing effect 
seriously.

\acknowledgments

I am grateful to S.J. Brodsky, F.A. Harris, J.L. Rosner, and 
S.F. Tuan for useful conversations and/or communications on the 
$\rho\pi$ puzzle and related subjects during the past few years. 
Thanks are also due to P. Wang for informing me of the prospect 
of the Beijing charm experiment. This work was supported in part 
by the Director, Office of Science, Office of High Energy and 
Nuclear Physics, Division of High Energy Physics, of the U.S.  
Department of Energy under contract DE-AC03-76SF00098 and in 
part by the National Science Foundation under grant PHY-0098840.


\begin{references}
\bibitem{comment}  The branching fractions of $\psi'$ quoted here
                   are mostly from Ref. \cite{Harris}. Those which
                   are not found in Ref. \cite{Harris} have been taken
                   from Ref. \cite{PDG}.
\end{references}
\end{document}